\documentclass{svmult}

\usepackage{graphicx}
\usepackage{makeidx}
\usepackage{multicol}

\begin{document}

\title*{Fast Spectroscopy And Imaging With The FORS2 HIT Mode}
\author{Kieran O'Brien}
\institute{European Southern Observatory, Santiago, Chile. \texttt{kobrien@eso.org}}
\maketitle

\begin{abstract}
The HIgh-Time resolution (HIT) mode of FORS2 has 3 sub-modes that allow for imaging and spectroscopy over a range of timescales from milliseconds up to seconds. It is the only high time resolution spectroscopy mode available on an 8~m class telescope. In imaging mode, it can be used to measure the pulse of pulsars and spinning white dwarfs in a variety of high throughput broad- and narrow-band filters. In spectroscopy mode it can take up to 10 spectra per second using a novel ``shift-and-wait'' clocking pattern for the CCD. It takes advantage of the user-designed masks which can be inserted into FORS2 to allow any two targets within the 6.8'~$\times$~6.8' field of view of FORS2 to be selected. A number of integration, or more precisely `wait', times are available, which together with the high throughput GRISMs can observe the entire optical spectrum on a range of timescales. 
\end{abstract}

\section{general description}

FORS is the FOcal Reducer and low dispersion Spectrograph for the Very Large Telescope (VLT) of the European Southern Observatory (ESO). It is designed as an all-dioptric instrument for the wavelength range from 330~nm to 1100~nm. It is capable of imaging as well as low to medium resolution spectroscopy, with a wide range of filters and grisms (see Tables~2.2, 2.4 and 2.5 of the FORS user manual\footnote{http://www.eso.org/instruments/fors/doc/} for a comprehensive list). Single and multiple object spectroscopy options are available using long slits, a set of 19 moveable jaws (referred to as MOS mode) and a magazine capable of holding up to 10 laser-cut Invar masks (referred to as MXU mode). 

FORS-2 saw first light in 1999 and, together with its non-identical twin FORS-1, has accounted for more refereed publications than any other instrument at the VLT. Following an upgrade in April 2002, FORS-2 has been equipped with a mosaic of two 2k~$\times$~4k MIT CCDs  (pixel size of 15~$\times$~15~$\mu$m) with a pixel scale of 0.125"/pixel, although it is operated with binning 2~$\times$~2 as standard delivering a final pixel scale of 0.25"/pixel. The HIT mode of FORS-2 was available from the beginning of operations~\cite{cumani1, cumani2}. However, it was subsequently unavailable for over a year following the upgrade of the CCD camera and has only recently been re-offered to the community. 

The general principle of the HIT mode is to increase the duty cycle of imaging and spectroscopic observations by reducing the exposed region of the CCD and shifting the charge from this small region to an unexposed `storage' region of the CCD. By shifting the charge in this manner a number of times, it is possible to store a time series of images or spectra on the CCD without needing to read-out the CCD after each exposure. The shutter remains open throughout the observation and is only closed once a predefined number of shifts have occurred. The CCD is then read-out using the standard low noise read-out mode used for all spectroscopic observations. Further details of the different clock patterns used can be found in Sects.~\ref{section:OSclockpattern}~\&~\ref{section:MSclockpattern}. 

The HIT mode is similar in concept to the (now decommissioned) Low Smear Drift (LSD) mode used by ISIS on the William Herschel Telescope on La Palma \cite{lsdmode}. This drift mode was limited by the size of the memory on the DMS (16~Mb) and allowed read-out of several spatial pixels in a given exposure. Another previous drift mode used the RGO spectrograph on the Anglo-Australian Telescope at Siding Spring, New South Wales \cite{rgomode}. This mode was similar to the LSD mode on ISIS, again using a physical buffer which limits the number of spectra that can be obtained in one exposure. In addition to these drift modes, a continuous readout mode was available as a visitor instrument on LRIS for a short time. This mode did not have the limitations of the buffer size and allowed a continuous read-out of the CCD. However, it did not allow for a comparison star and suffered from several instrumental effects that were never overcome (see \cite{ksothesis})

\section{HIT modes}

There are three modes available to users that offer very different characteristics and are suitable for a wide variety of applications.

\begin{itemize}
\item HIT-I: Imaging mode
\item HIT-OS: One-shift spectroscopy mode
\item HIT-MS: Multiple-shift spectroscopy mode
\end{itemize}

\subsection{HIT-I}
\label{section:OSclockpattern}

The HIT-I mode is the only available imaging mode. It uses the moveable slitlets of the MOS unit located in the top section of the instrument to form a pseudo-longslit by aligning them along the same column or columns. The slit width can be set to a value in the range 0.2--30" and is placed so that the projection of the slit falls on the row nearest to the serial register in the unvignetted region of the field of view. 

\begin{figure}
\centering
\includegraphics[height=9cm, angle=0]{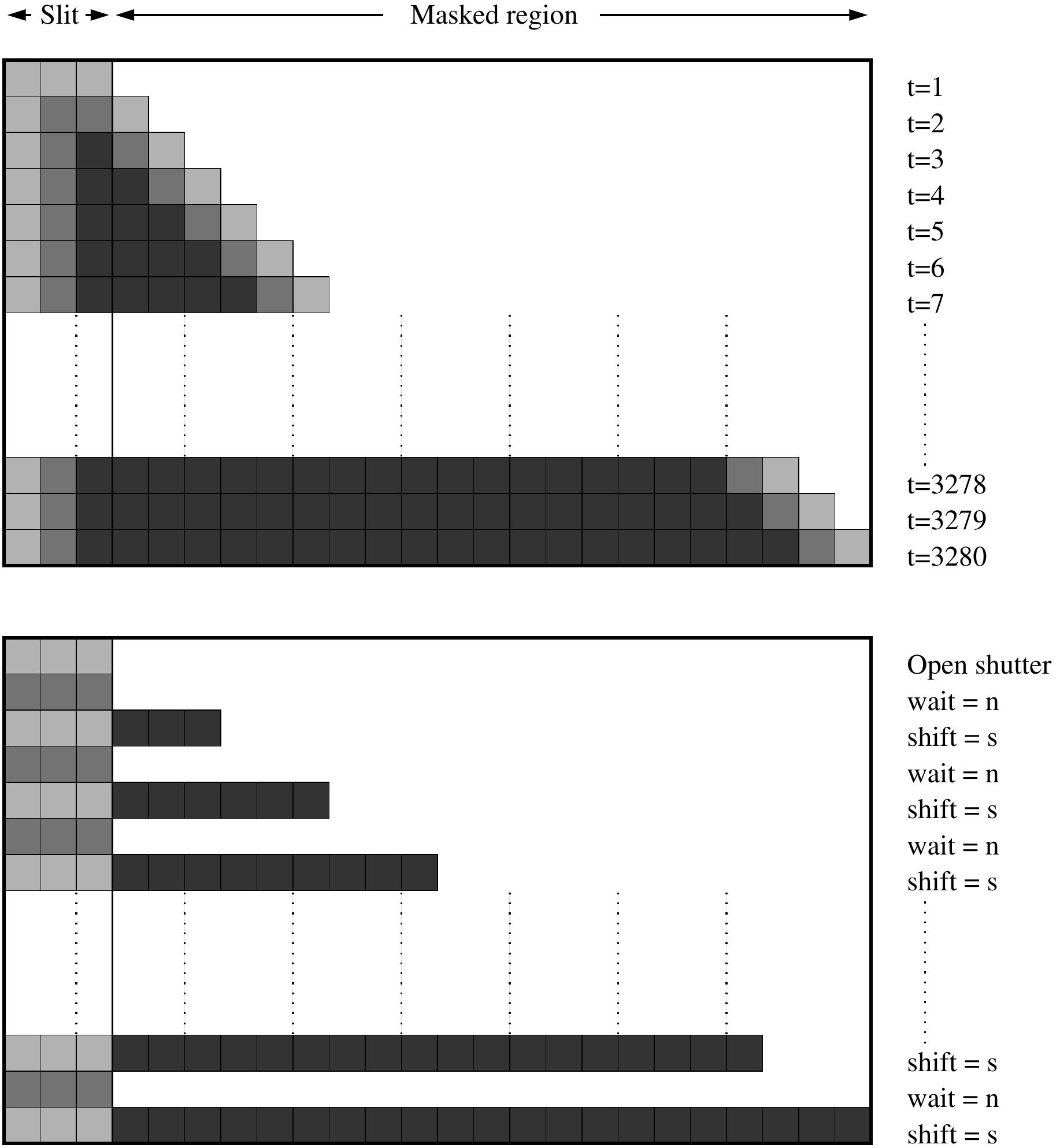}
\caption{The basic operation of the different HIT mode clock patterns: {\bf (a)} the top panel shows the clock pattern for the HIT-I and HIT-OS patterns, where each row is shifted individually until the first row shifted reaches the end of the unvignetted region; {\bf (b)} the bottom panel shows the clock pattern for the HIT-MS mode, where a number of rows are grouped together and shifted out from the exposed region into the masked region where they wait for a time while the pixels under the slit are exposed to light from the source. `s' refers to the size in unbinned pixels of the slit, whilst `n' is the number of seconds to wait between shifts }
\label{figure:clockpatterns}  
\end{figure}

The basic scheme of the HIT-I clock pattern is shown in the top panel of Fig.~\ref{figure:clockpatterns}. The imaging mode is available with five variations on this clock pattern. In each the charge is shifted across the unvignetted region of the chip (3280 unbinned pixels) in the direction away from the serial register, but the speed at which the charge is moved is varied to simulate a longer or shorter exposure time. The transfer from t=1--3280 takes place in either 1, 4, 16, 64 or 256 seconds, which identifies the clock pattern (i.e. HIT-OS1-1sec ... HIT-OS5-256sec). This means each row is shifted every 0.61, 2.4, 9.8, 39.0 or 156~milliseconds. The speed of the slowest mode is determined by technical limitations. It is similar to the shortest exposure time for full frame imaging available with FORS-2 (0.25~seconds). Once the final shift has taken place, the shutter is closed and the frame is read-out in the normal manner, using the low read-out noise mode of FORS, used for spectroscopic observations (100khz,2x2,high).

The time resolution of the resulting lightcurve depends on the slit width, as each pixel is exposed as long as it remains in the unmasked region of the CCD. As can be seen in Fig.~\ref{figure:clockpatterns} there are also some edge effects if the image of the slit is projected onto more than one pixel.

All of the filters available in the FORS filter-set can be used, which includes standard Bessel filters, as well as Gunn and an number of narrow-band filters centred on common emission lines (H$\alpha$, HeI, OII). In addition it is possible for users to supply their own filters (details of this process can be found on the FORS webpages). The slit created by the MOS, which covers the entire 6.8' field, can be rotated on the sky to include a comparison star for relative photometry.

The limiting magnitude for the HIT-I mode is difficult to quantify as there are several parameters that have a large influence on the limit. In addition, many of these also have direct influence on other characteristics of the final data-set. Here I will summarize the most important of these,

\begin{itemize}

\item {\bf Clock pattern:} The clock pattern determines the amount of time the pixels are exposed to light and can be thought of in the same way as the exposure time in traditional imaging.

\item {\bf Slit width:} Increasing the slit width will let more light onto the chip and therefore allow fainter objects to be observed. However, it will also increase the projected area of the slit on the CCD and mean that each pixel is illuminated for more than one clock-cycle, thus reducing the time resolution of the resulting lightcurve by convolving the true lightcurve with a filter with a width equal to the number of pixels exposed (and hence the number of time steps). 

\item {\bf Filter:} Obviously, the filter transmission has a direct influence on the limiting magnitude.

\item {\bf Atmospheric conditions:} Atmospheric conditions such as seeing and transparency will lead to larger slit losses. 
\end{itemize}

The magnitude limits in Tab.~\ref{table:maglimits} correspond to a S/N of 5 per time step using the most popular mode, HIT-OS4-16sec. They have been calculated for a slit-width of 0.25'' (equivalent to 1 pixel), which results in a time resolution of 9.75~milliseconds. They are calculated assuming a dark sky, clear conditions, a seeing FWHM of 0.8'' and an airmass of 1.2, and have been determined for a point source of zero colour (A0V star). 

\begin{table}
\begin{center}
\begin{tabular}{cc}
Filter & Magnitude limit \\
U & 12.6 \\
B & 16.5 \\
V & 17.1 \\
R & 17.3 \\
I &  16.7 \\
\end{tabular}
\caption{Magnitude limits for the HIT-I mode. These magnitude limits have been determined using the HIT-OS4-16sec mode with a slit-width of 0.25'', assuming dark sky and typical conditions for Paranal}
\label{table:maglimits}
\end{center}
\end{table}

Obviously, fainter targets can be observed by increasing the slit width and/or changing the readout mode. However, this increase is at the expense of the time resolution (due to increasing the number of pixels the slit is projected onto) and the photometric accuracy (due to the increased chance to image motion effects within the slit, see Section~\ref{section:imagemotion}).

\subsubsection{HIT-I observations of the Crab pulsar}

\begin{figure}
\centering
\includegraphics[height=7cm, angle=0]{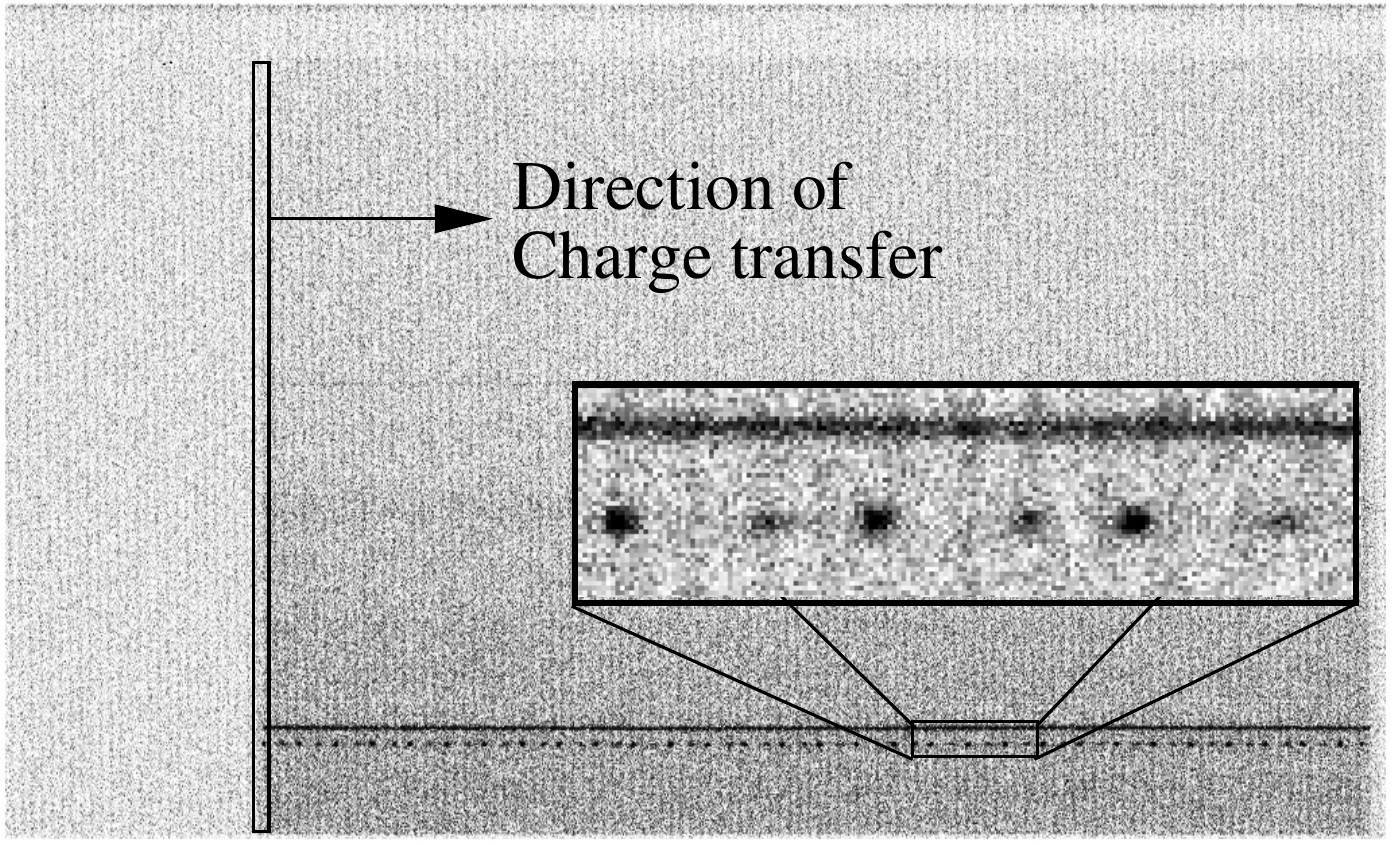}
\caption{HIT-I image of the Crab pulsar. The position of the slit is shown on the left hand-side of the field, which represents the exposed region of the CCD. Charge is shifted in the direction shown. The enlarged region shows both the crab pulsar (lower), including the major and minor peaks, as well as a nearby comparison star (upper). Time runs in the opposite direction to the charge transfer}
\label{fig:crab}  
\end{figure}

As a test of the timing accuracy we performed observations of the Crab pulsar. An example of the resulting dataset can be seen in Fig.~\ref{fig:crab}, which clearly shows the quality of the data that is produced by the HIT-I mode. In Fig.~\ref{fig:crab}, the position of the slit is clearly marked on the left hand-side of the image together with a further marker indicating the direction the charge is shifted across the CCD. A section of the image has been enlarged to show the trace of the pulsar (bottom) together with that of a nearby bright comparison star to show what can be expected from this mode. The ``major peak'' can be clearly seen, as well as the fainter ``minor peak''. The minor peak occurs slightly before $\phi_{pulse}=0.5$ in the 33~millisecond pulse period (note that since the charge is shifted from left to right, the first images will appear on the right-hand side of the final image. These images were taken with the HIT-OS1-1sec mode with a slit width of 5'' under average conditions (an airmass of 1.5, FWHM of the seeing of 0.8", dark sky) through the R\_SPECIAL filter.

Whilst the HIT-I mode of FORS2 is relatively unique in offering the capability to perform imaging on a sub-second timescale using a CCD detector, it remains inferior to more dedicated HTRA instruments in a number of ways. For instance, ULTRACAM\footnote{See the contribution elsewhere in these proceedings for a description of this instrument} has the advantage of having three colours simultaneously allowing colour changes to be determined, as well as a much larger field-of-view, which allows for a range of comparison stars to be selected, leading to much better determined photometry. SALTICAM\footnotemark[\value{footnote}]  again has a large field-of-view from with to select comparison stars for relative photometry, but is limited in the amount of time it can observe a give target for. OPTIMA\footnotemark[\value{footnote}]  and other similar photon-counting devices are able to observe a range of different phenomena on even faster timescales, but suffer from low throughput. In addition, the HIT mode is permanently mounted on an 8.2~m telescope (the VLT) and is available in service mode, for observations on very short notice, with a wide range of filters. The VLT allows for much longer observations than are possible with SALT and can react very quickly to transient phenomena over a large fraction of the sky.
 
\subsection{HIT-OS}

The HIT-OS mode was the first high time resolution spectroscopy mode available at the VLT. It uses masks cut with the laser cutting machine of the VIMOS instrument. As the user is limited to single target spectroscopy, there are a number of pre-cut masks for the HIT-OS mode with square apertures of varying sizes (0.5--5.0''). It is only possible to put a single target in the aperture cut in the mask as the charge would otherwise be shifted under the dispersed light of the subsequent apertures. 

Due to the fact that the standard dispersion direction is perpendicular to the serial register which is incompatible with the direction in which the charge is shifteded in the HIT modes, only 2 grisms, which have been rotated by 90$^{\circ}$, are available for the HIT-OS mode. The characteristics of these are shown in Tab.~\ref{table:grisms}.

\begin{table}[h]
\begin{center}
\begin{tabular}{cccc}
Grism name & central wavelength & dispersion & limiting magnitude \\ 
& (nm) & (nm/pixel) & \\
600B & 445 & 0.075 & 15.8\\
300I & 857 & 0.162 & 15.9 \\ 
\end{tabular}
\caption{Details of the grisms available for the HIT-OS and HIT-MS modes, together with the limiting magnitude for the HIT-OS mode}
\label{table:grisms}
\end{center}
\end{table}

The HIT-OS mode uses the same clock patterns as the HIT-I mode and is therefore limited in the number of targets that can be observed due to the relatively high limiting magnitude, shown in Tab.~\ref{table:grisms}. The limiting magnitudes have been calculated using the HIT-OS5-256sec clock pattern, which results in a time resolution of 0.6*(slitwidth in pixels). However, if you want to reach this limit and ensure some degree of photometry, it is necessary to use a mask with a large aperture in order to minimize the slit losses. This photometric stability comes at the cost of the spectral resolution (which will be limited by the seeing profile parallel to the slit), the spectral stability (due to image motion parallel to the dispersion direction), the temporal resolution (which will be limited by the seeing profile perpendicular to the slit) and the short time-scale photometric stability (which will be limited by image motion perpendicular to the slit). However, if you choose a narrow, short slit (i.e. a small aperture) then you will only be able to measure equivalent width and line profile changes. The stability issues and the relatively bright limiting magnitude greatly limit the applicability of the mode, which is why we have introduced a second spectroscopic mode, the HIT-MS mode which overcomes many of these problems.

\subsection{HIT-MS}

The HIT-MS mode can be thought of as ``\,`normal' spectroscopy with large gains in duty cycle'' as it allows the user to operate in a pseudo-MOS mode, storing multiple exposures of the same target (or targets) on the CCD between successive read-outs of the CCD.

\subsubsection{Multiple shift clock pattern}
\label{section:MSclockpattern}

The HIT-MS mode allows users to operate the CCD with a ``shift-and-wait'' clock pattern, as shown in the bottom panel of Fig.~\ref{figure:clockpatterns}. In contrast to the HIT-I and HIT-OS modes, where the charge is constantly being shifted from one row to the next and only the rate at which it is shifted is changed, the HIT-MS clock pattern shifts a pre-defined number of rows very fast ($\sim$2.5~microseconds per line) and then integrates for a pre-defined `wait' time before the sequence is repeated. This continues until the maximum number of line-shifts has occurred and the CCD is read-out using the standard low read-out noise spectroscopic mode (100kHz,2x2,high). 

\subsubsection{Mask design}

The HIT-MS mode again uses masks inserted into the MXU of FORS2 and is thus limited to FORS2. However, in contrast to the HIT-OS mode, the user designs the masks, as it is possible to have a number of slitlets and hence targets. There is no real limit to the number of slitlets, as long as the number of rows shifted each time is equal to the number of rows between the projection of the top of the uppermost slit to the bottom of the lowermost slit. In practice, it is usually necessary to have just two slits (one for the target and one for a nearby comparison star to be used to correct for slit-losses, small instrumental wavelength shifts and seeing variations). With just two slitlets it is possible to rotate the instrument so that the projection of the bottom of the uppermost slit falls on the row above the top of the lowermost slit, thus minimizing the number of rows that are shifted each time. 

\begin{figure}[htbp]
\begin{center}
\includegraphics[height=8cm, angle=0]{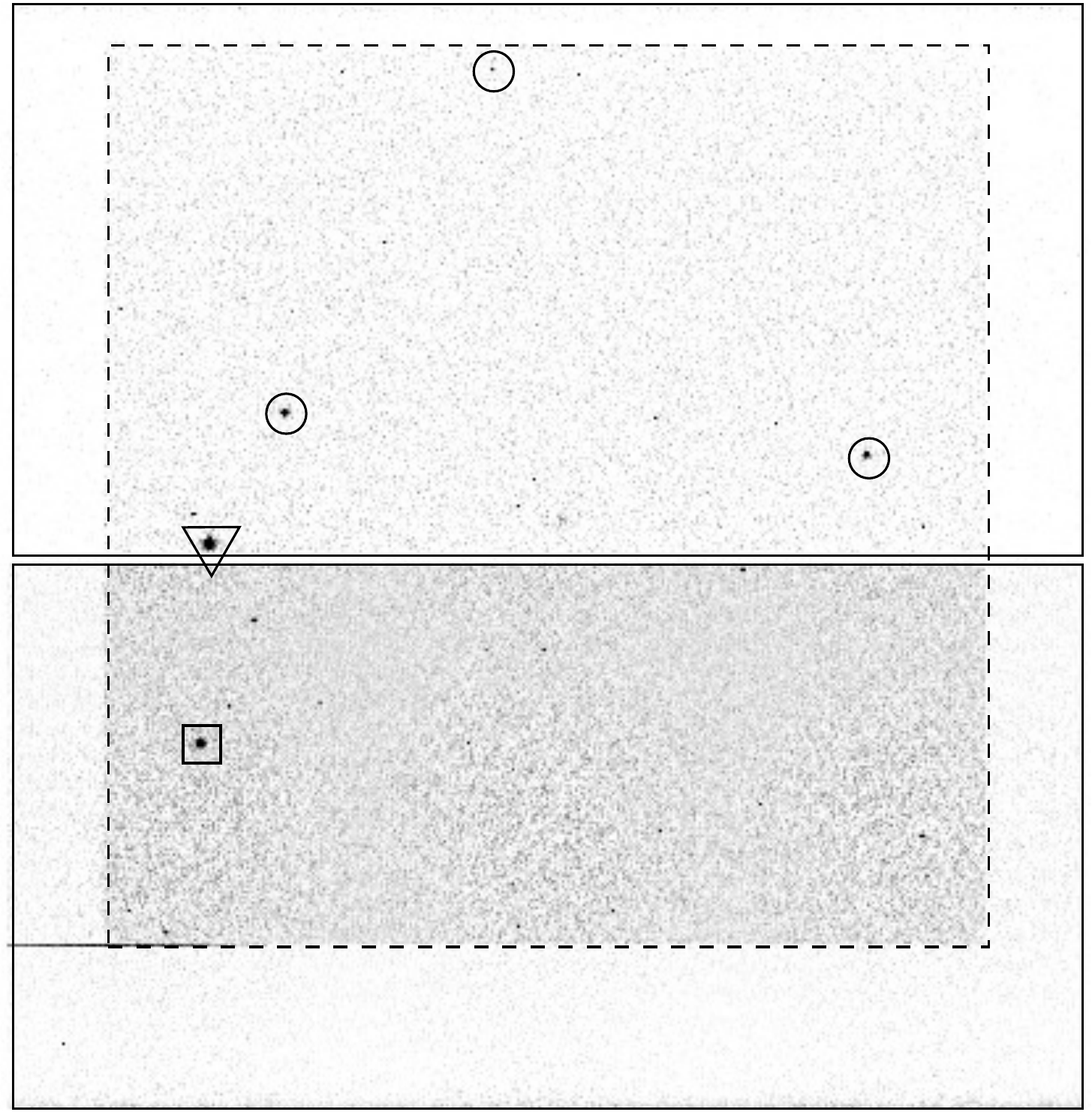} 
\caption{A typical acquisition image showing the relative orientations of the upper and lower CCDs ({\it solid rectangles}), the field of view of the instrument ({\it dashed square}), the alignment stars ({\it circles}), the comparison star ({\it triangle}) and the target ({\it small square})}
\label{fig:acq}
\end{center}
\end{figure}

A sample acquisition image is shown in Fig.~\ref{fig:acq} which highlights all of the important features. Towards the top of the image are the alignment stars ({\it circles}) which the software (based on a number of MIDAS routines) uses to determine the small-scale offsets of the mask with respect to their nominal positions. Once these offsets have been calculated they are applied to the telescope before the mask is inserted in order to ensure the optimal alignment on sky. The position of the target ({\it small square}) is shown, which, in this case, is offset from the centre of the field in order to maximise the spectral coverage of the resulting spectrum. As dispersion in the the y-direction it is important to ensure that the gap in the CCD covers a relatively uninteresting part of the spectrum. The position of the comparison star ({\it triangle}) which has been orientated so that it is 5'' away from the target in the x-direction, ensuring that the images from the slit do not overlap, as described above. 

One consequence of the offset between the target and comparison star slits is that the central wavelength of the resulting spectra will not be the same (due to their offset in the y-direction, the direction of the dispersion). This will make the correction for slit losses slightly more complicated, as the wavelength coverage is not the same for both targets, as would be the case in long-slit spectroscopy. Up to 10 masks can be inserted into the instrument during daytime in order to allow for either a number of targets, or a range of slit-widths to counter the problem of changing atmospheric conditions during the night. In contrast to the HIT-OS mode, narrow slit-widths are possible without the complete loss of photometric accuracy due to the presence of a comparison star, whose purpose it is to correct for such effects. This removes the problems caused by motion across the slit when the star under-fills the slit. In addition, accurate sky extraction is possible from regions around both the target and the comparison stars.

Both the target and comparison stars are moved to the left-hand side of the field to allow far as many shifts as possible before the CCD needs to be readout. 

\subsubsection{Duty cycle gains}

The number of spectra that can be stored on the CCD before it is read out depends on the number of pixels shifted per `shift-and-wait' cycle. As described in the previous section, the user would typically define two slitlets, one each for a target and a nearby comparison star. Assuming each slitlet is 5'' long, which is long enough to sample the sky region around the star under typical conditions, then the number of rows shifted each step would be 80 (2 slitlets * 5'' * 8 pixels/''). The maximum rows available is 3280, meaning a maximum of 41 shifts before the CCD needs to be read out. 

\begin{table}[h]
\begin{center}
\begin{tabular}{lccccc}
clock pattern & wait time & cycle time & HIT duty cycle & `normal' duty cycle & gain factor \\ 
& (secs) & (secs) & (\%) & (\%) & \\ 
HIT-MS1-01sec & 0.1& 44.1 & 9.3 & 0.2 & 46 \\
HIT-MS2-02sec & 0.2& 48.2 & 17 & 0.5 & 34 \\
HIT-MS3-05sec & 0.5& 60.5 & 34 & 1.2 & 28 \\
HIT-MS4-1sec & 1.0& 81.0 & 51 & 2.4 & 21 \\
HIT-MS5-2sec & 2.0& 122 & 67 & 4.8 & 14 \\
HIT-MS6-5sec & 5.0& 245 & 84 & 11.1 & 7.6 \\
HIT-MS7-10sec & 10.0& 450 & 91 & 20.0 & 4.5 \\
HIT-MS8-20sec & 20.0& 860 & 95 & 33.3 & 2.9 \\
\end{tabular}
\caption{The duty cycles for the currently d HIT-MS clock patterns. The overheads per read-out include read-out of the CCD ($\sim$10~seconds) and set-up time ($\sim$30~seconds). These are not significantly improved by windowing of the CCD}
\label{table:dutycycles}
\end{center}
\end{table}

Table~\ref{table:dutycycles} shows the duty cycles for `normal' and HIT mode spectroscopy together with the currently offered MS clock patterns. As can be seen from this table, for short integration times, the HIT mode is almost fifty times more efficient than using the traditional `single exposure + readout' pattern. Even for wait times as long as 20~seconds, the HIT mode remains 3 times more efficient. In addition to these gains in efficiency, the HIT mode does not suffer from long deadtimes between exposures, making it possible to probe the variability on timescales of the order of the exposure (or `wait') time, which is not possible with other (more traditional) modes of operation. 

\section{Characteristics of the HIT sub-modes}

\subsection{Timing accuracy}

The time signal of the observatory clock is distributed in two ways; firstly, via Network Time Protocol (NTP) for systems needing an internal accuracy of $\sim10$~milliseconds and secondly, via dedicated timing (TIM) boards that deliver an internal accuracy of microseconds. These times are synchronised to a rubidium clock that is in turn synchronised annually via a GPS receiver. This system is expected to give an absolute accuracy on the timestamp of around $1$~millisecond or better. The timestamp is placed on each frame and represents the start time of the exposure (in fact it is the time when the shutter has just begun to open, which is when the shifting starts). In order to calculate the time for a given pixel it is also necessary to know how many shifts have taken place.

\begin{figure}[htbp]
\begin{center}
\begin{tabular}{cc}
\includegraphics[height=6cm, angle=270]{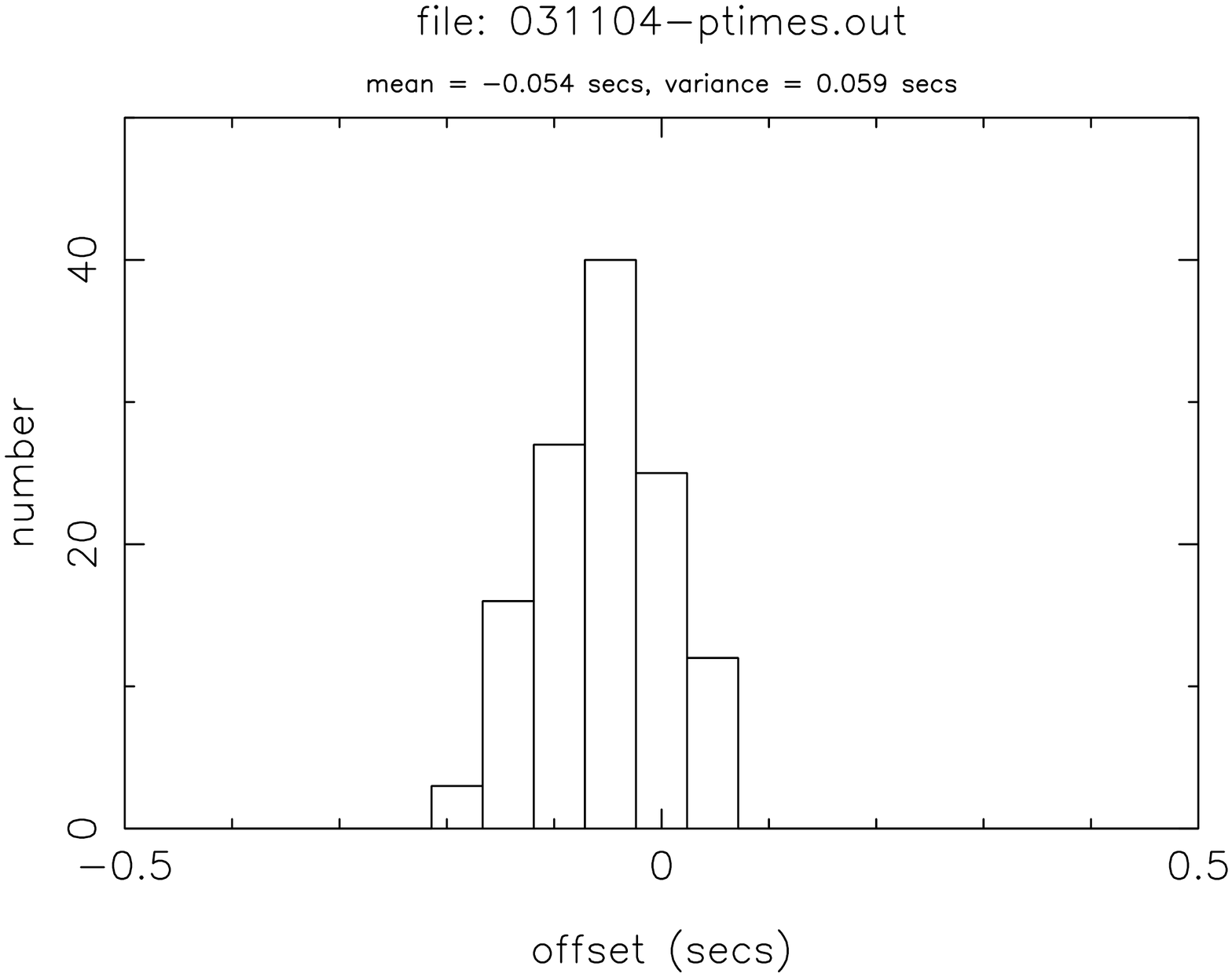} & 
\includegraphics[height=6cm, angle=270]{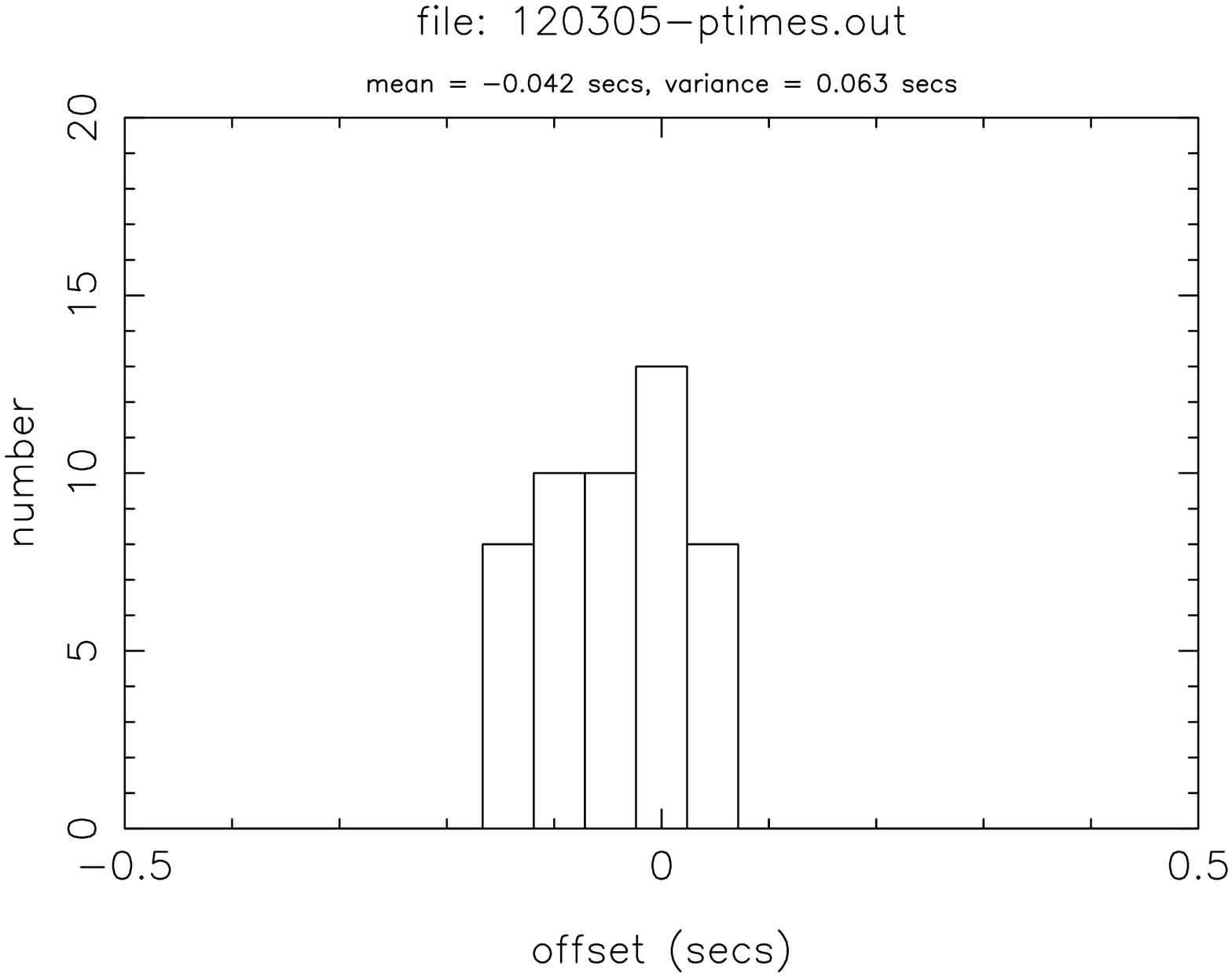} \\\includegraphics[height=6cm, angle=270]{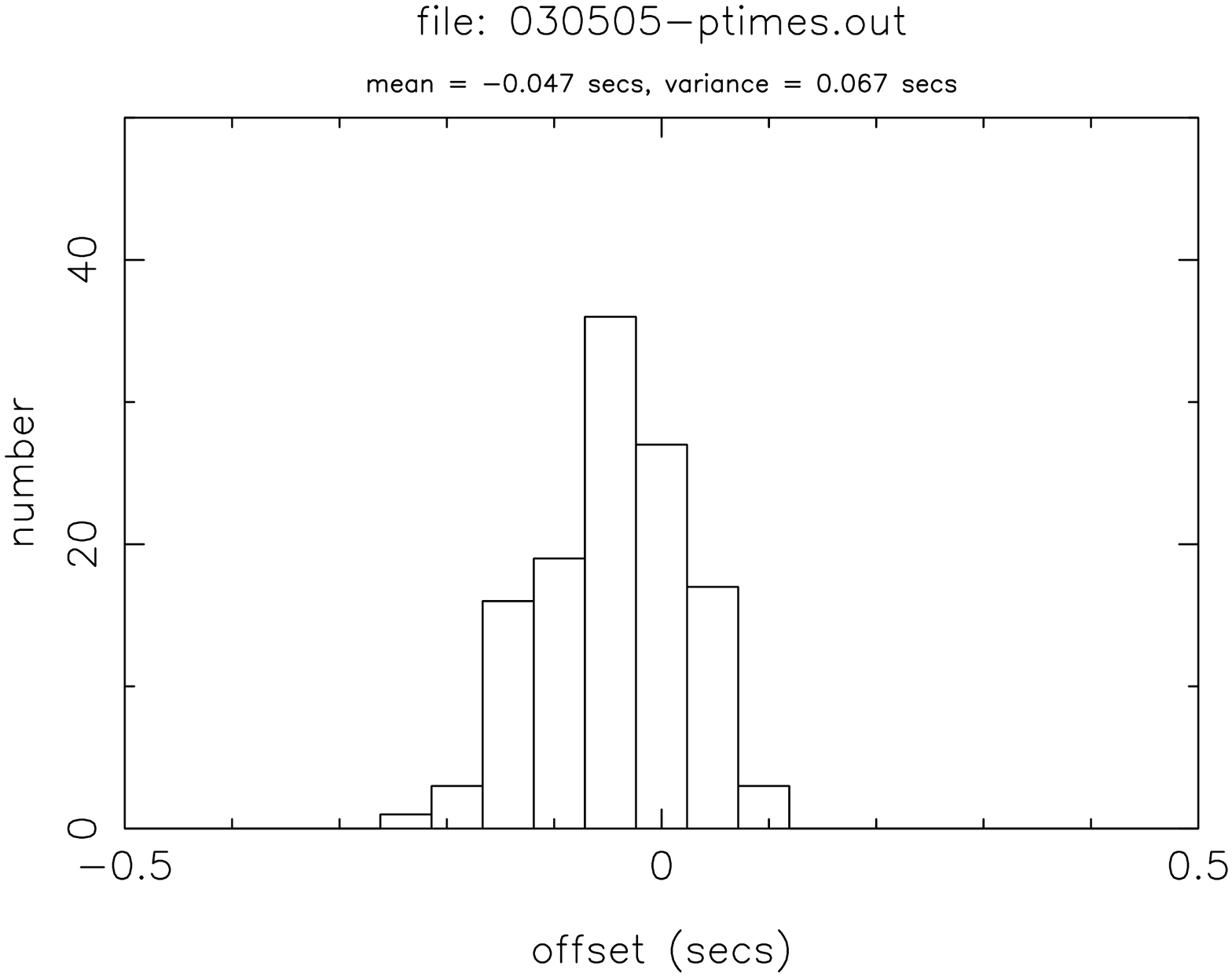} & \\
\end{tabular}
\caption{Histograms of the timing marks used to determine the absolute timing offset of the data and the variance in that value. The data were taken on: {\bf (a)} 3rd November 2004 (top left); {\bf (b)} 12th March 2005 (top right); {\bf (c)} 3rd May 2005 (bottom left)}
\label{fig:timestamp}
\end{center}
\end{figure}

In order to determine the timing accuracy of the HIT mode, we illuminate the CCD using a small bundle of LEDs that are set to trigger on the 1 pulse-per-second signal from the TIM board and last for 200~milliseconds. We then record a number of frames with the HIT-OS1-1sec mode and determine the position of the leading edge of the pulse. If we assume that  our accuracy is better than one second, then any residuals from a integer value allow us to determine the accuracy of the time-stamps relative to the observatory clock and, in turn, UTC. The results of 3 such runs are shown in Fig.~\ref{fig:timestamp}. As can be seen, the absolute timing accuracy (as given by the mean offset) and the relative timing accuracy (i.e. one stamp relative to another, as given by the square root of the variance) are $\sim50\,\&\,7.7$~milliseconds respectively. This means that there is an instrumental delay of 50~milliseconds that can safely be removed as an offset for a given set of observations (and can be calibrated before an observing run). However, the 7.7~millisecond relative delay is the `jitter' between frames that cannot be removed. This means that between two frames (or, equivalently, lightcurves) there is a random offset of $\sim$11~milliseconds.

\subsection{CCD parameters \& noise model for the HIT mode}

The FORS2 detector consists of two 2k~$\times$~4k MIT CCDs (15 $\mu$m pixel size). The MIT CCDs were optimised to provide much higher response in the red wavelength range beyond 800~nm, with impressively low fringe amplitudes. However, this was at the expense of the response below 400~nm, which makes them sub-optimal for many HTRA applications. The standard CCD parameters are summarised in Tab.~\ref{table:ccdparameters}. As stated previously, the CCD is operated with 2~$\times$~2 binning as standard. 

\begin{table} [h]
\begin{center} 
\begin{tabular}{lccc}
Parameter & unit & Chip 1 & Chip 2 \\
read-out noise & ADU & 2.7 & 3.0 \\
gain & e$^{-1}$/ADU & 0.7 & 0.7 \\
Dark current @-120$^{\circ}$C & e$^{-1}$/pix/hour & $\sim3$ & $\sim3$ \\
linearity & \% RMS & 0.1 & 0.23 \\
readout time & seconds & \multicolumn{2}{c}{41} \\
\end{tabular}
\caption{CCD parameters for the `100kHz,2$\times$2,high' readout mode of the FORS2 detector; source FORS user manual and ODT pages}
\label{table:ccdparameters}
\end{center} 
\end{table} 

The noise model for the HIT mode is the same as that for the standard spectroscopic readout mode. The only additional source of noise is simply due to the longer dark time for the exposure. In the case of the longest mode (HIT-MS8-20sec) the charge remains on the chip for 1000~seconds, so will typically incur a penalty of $\sim$1~e$^{-}$/pix.
 
\subsection{Image motion}
\label{section:imagemotion}

Image motion within the slit is a major source of uncertainty for the photometric accuracy of the HIT-I and HIT-OS modes, but not necessarily for the HIT-MS mode, as it is possible to use narrow slit widths. If the direction of motion is along the slit, then the variability is simply seen as image motion. However, if the image motion is across the slit, then the effect is seen as a series of maxima (at times when the image motion is in the same direction as the charge is being shifted) and minima (when the motion is in the opposite direction) in the lightcurve. This can be seen in the example below where a standard star was placed in a 5'' wide slit using the HIT-OS1-1sec mode. 

\begin{figure}[htbp]
\begin{center}
\includegraphics[height=11cm, angle=270]{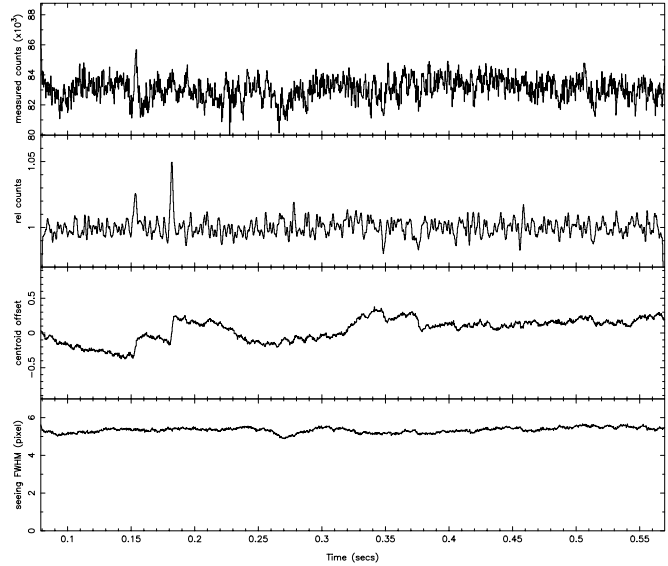} 
\caption{The effects of image motion and seeing changes on the HIT-I and HIT-OS lightcurves. The upper panel shows the flux lightcurve extracted by fitting the seeing profile from an observation of a photometric standard star. The upper-middle panel shows the model lightcurve using the parameters from the fit (as described in the text). The lower-middle and bottom panels show the position of the centroid (in pixels relative to the first point) and the FHWM of the measured seeing profile respectively.  The image scale is 0.25~"/pixel}
\label{fig:gaussfit}
\end{center}
\end{figure}

In order to study the effects of the image motion on the lightcurves, we have fit the profile of the star at each time step using a minimum chi-squared goodness of fit approach. The model profile used was a Gaussian with the peak, centroid and width as free parameters. The best fit values for each of these can be seen in the top, lower-middle and lower panels of Fig.~\ref{fig:gaussfit}. 

As the centroid has been derived from the 1-D profile of the star along the slit, it is fair to assume that this is also a plausible model for the motion in the direction {\bf across} the slit, which is otherwise impossible to determine. Likewise, it is also reasonable to assume that the seeing varies symmetrically, so that the variability of the FWHM of the seeing (or more accurately the measured image quality) is qualitatively the same in both directions. For these reasons, we have used these parameters to investigate the effects of image motion and seeing on the observed flux from a non-variable source, eg. a standard star. The model lightcurve is created from the parameters described above by projecting the `star' through a `slit' and calculating the expected flux at each time-interval/shift. The resulting lightcurve for a 5'' slit width (the same as in the real observations) is shown in the upper-middle panel of Fig.~\ref{fig:gaussfit}. There are a number of interesting features in this lightcurve;

\begin{itemize}

\item t=0.155~seconds: This feature is present in both the observed and the model lightcurves and seems to be caused by the rapid image motion within the slit. This indicates that the motion took place both along and across the slit. 

\item t=0.18~seconds: This feature is only present in the model lightcurve and again seems to be due to the image motion. However, as it is not present in the observed lightcurve, we can infer that the motion was only along the slit. 

\item t=0.27~seconds: The dip in the observed lightcurve is associated with a slight improvement in the 1-D seeing. However, such an improvement should not have any effect on the lightcurve, due to the large slit width. We can infer from this that either the seeing degraded significantly in the direction across the slit, leading to slit losses, or the transparency changed briefly.

\end{itemize} 

\begin{figure}[h]
\begin{center}
\includegraphics[height=11cm, angle=270]{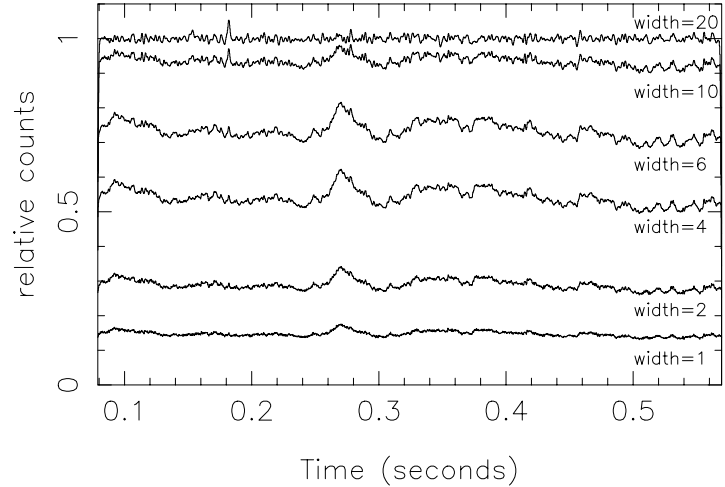} 
\caption{The effect of the slitwidth on the model lightcurves. The lightcurves have been calculated using a range of slitwidths from 1--20 binned pixels, which corresponds to 0.25"--5'' }
\label{fig:slitwidth}
\end{center}
\end{figure}

The effects of the slit width can be seen in Fig.~\ref{fig:slitwidth}, where the parameters from the fitting mentioned above are again used to simulate a lightcurve. This time the slitwidth in pixels is varied from 1 to 20 pixels (0.25" to 5''). When the slitwidth is 20 pixels no major deviations from unity are seen, indicating no slit losses. However, even with a slitwidth of 10~pixels some losses are seen except in the case where the seeing is at its best. This trend continues until the slitwidth becomes comparable to the seeing when it again becomes rather insensitive to the seeing (and image motion), but the average value is by this time far from unity, indicating a large slit-loss.

\section{FO Aqr - a worked example}

FO Aqr belongs to the sub-class of Cataclysmic Variables (CVs) called Intermediate Polars (IPs). IPs contain a magnetic white dwarf which is accreting material from a low-mass companion. FO Aqr has a 291~minute orbital period, which is clearly seen in photometry and spectroscopy of the system. In addition to this there is a clear 20.9~minute `pulsation' period that is caused by the spin of the white dwarf, that has been spun up to these speeds by accretion torques. The inner accretion disk is truncated by the magnetic field of the spinning white dwarf which causes a modulation of the optical (and X-ray) flux on the spin and spin-orbital beat frequencies. This range of phenomena made FO Aqr an ideal target for highlighting the capabilities of fast spectroscopy on a complex but relatively well understood system (see e.g. \cite{hellier90, marsh96}).

We observed FO Aqr between UT07:42 - 10:37, 07 August 2005 using the HIT-MS mode with a `wait' time of 2.5~seconds. This resulted in a total of 3224 spectra of FO Aqr. We observed a simultaneous comparison star which was used to correct for slit losses and slight wavelength drifts due to high airmass of the source at the time of observation. We used the 600B grism and covered the range 380--675~nm. A slit width of 1'' was used resulting in a resolution of 780 at the central wavelength of 465~nm. 

\subsection{Data reduction}
The steps needed to reduce HIT-MS mode data are similar to those needed for any other spectroscopic observation. Each image can be thought of as a MOS image, with identical pairs of spectra taken during each wait time. A simple recipe for the data reduction based loosely around the steps used in PAMELA and MOLLY reduction software\footnote{Use of the MOLLY and PAMELA software developed by T. R. Marsh is gratefully acknowledged.} would be;

\begin{itemize}
\item {\bf Bias subtract} - The CCD is read-out in a completely standard way once the charge transfer steps have been completed and the shutter closed, so standard bias frames can be used.
\item {\bf Flatfield} - The flat field frame for the HIT-MS mode is different to other modes, as the effects of charge traps, etc. are spread over a number of rows due to the charge transfer steps. Special screen flatfield frames are taken during the daytime with the telescope at zenith.
\item {\bf Target location} - The centroid of the spatial profile of the target and comparison star are determined by fitting a Gaussian profile to the collapsed 1-D spatial profile
\item {\bf Trace dispersion} - the tilt of the spectrum (i.e. the displacement of the centroid of the spatial profile along the columns) is traced to ensured that the sky spectrum produced is the correct one. The tilt on the spectrum can be large between the two extremes of the spectrum ($\sim$ 20 pixels) and must be taken into account.
\item {\bf Sky spectrum} - a 2-D ÔimageÕ of the sky is created from regions sufficiently far away from the stellar profiles. This process also takes into account the tilt of the spectrum. 
\item {\bf Spectral extraction} - The spectra of the target and the comparison star are extracted using either normal or optimal extraction routines that use information from the spectral profile, the trace and sky images.
\item {\bf Wavelength calibration} - The spectra are calibrated in wavelength using arc lamp spectra taken during the daytime with the telescope at zenith. Further calibration can be performed using night sky lines and/or any features in the spectrum of the comparison star.
\item {\bf Slit losses} - The relative slit losses, due to the changing conditions, can be compensated for by fitting a low order polynomial to the comparison star spectrum and comparing these to the flux level in the average spectrum. In addition, if a spectrum of the comparison has been taken through a wide slit in photometric conditions, it is possible to calibrate the spectra on an absolute scale. As there will almost certainly be a discrepancy between the wavelength coverage of the target and the comparison star, it will be necessary to extrapolate the polynomial fit to cover the entire spectral range of the target star. As this is a smoothly varying function over the possible wavelength ranges, this should still give a satisfactory correction.
\item {\bf Flux calibration} - If a spectro-photometric standard star was observed during the night, it is also possible to extract its spectrum using the same calibration steps and determine the absolute flux scale. As in the previous step, the standard star may not have the same wavelength range as the target star and the calibration will need to be extrapolated.
\end{itemize}

%\begin{figure}[htbp]
%\begin{center}
%\includegraphics[height=10cm, angle=0]{dataframe.pdf} 
%\caption{The 2-D image of a sample CCD frame. The spectra can be seem tracing vertical lines and alternating from left to right with target and comparison star spectra. Several lines, most prominently H-alpha at the upper edge of the image can be seen in the 2-D image.}
%\label{fig:exampleframe}
%\end{center}
%\end{figure}

\begin{figure}[htbp]
\begin{center}
\begin{tabular}{cc}
\includegraphics[height=6cm, angle=270]{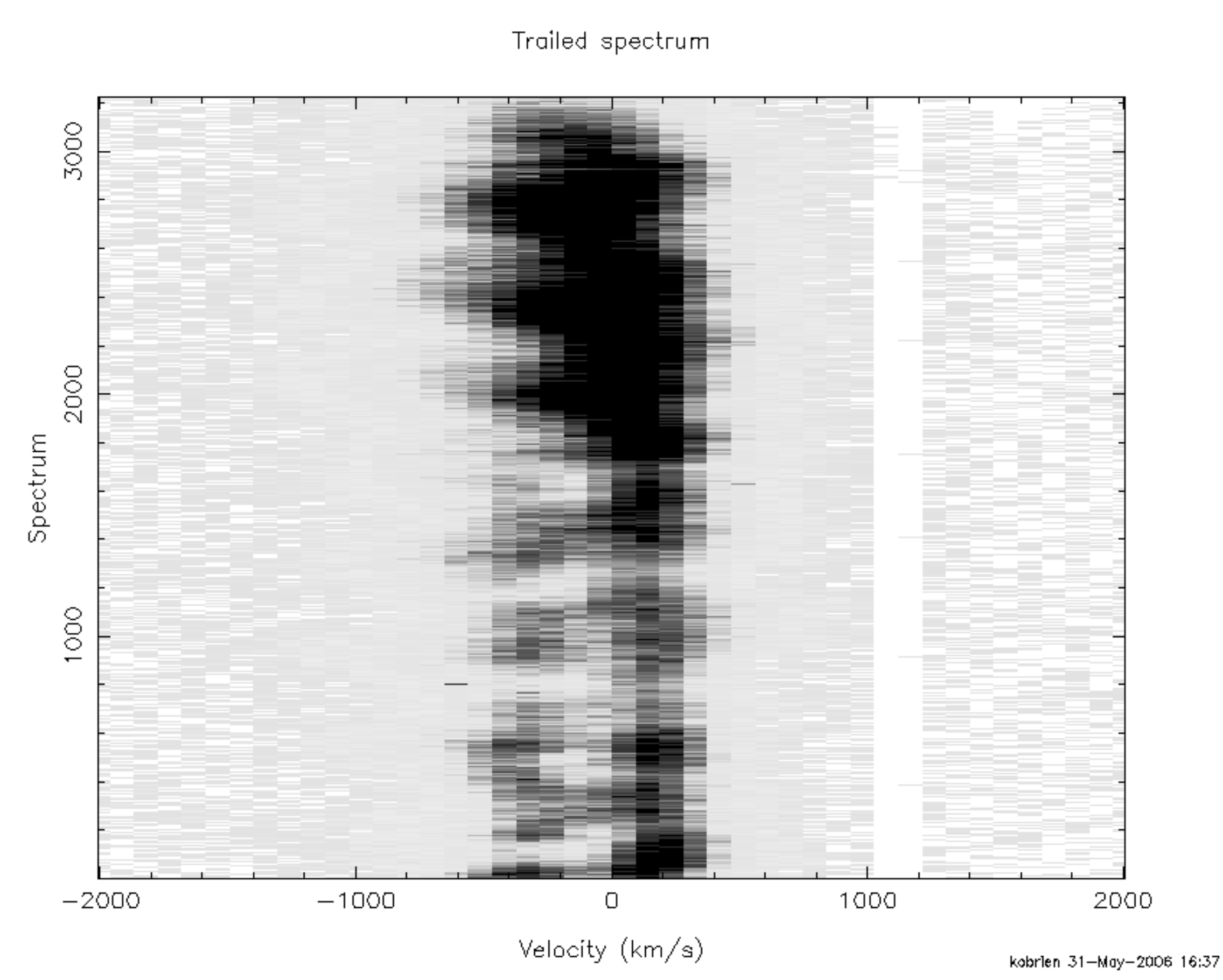} & \includegraphics[height=6cm, angle=270]{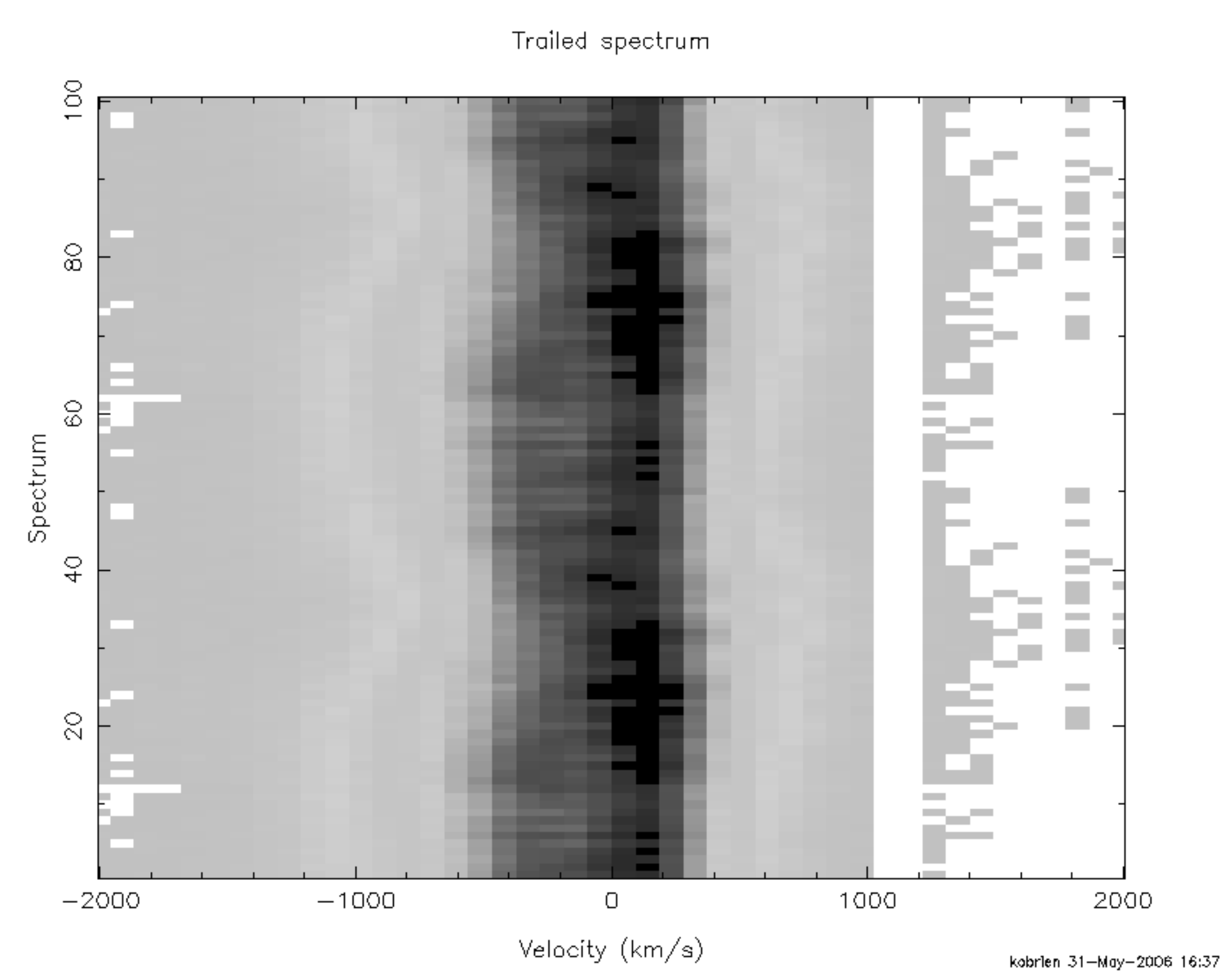} \\
\includegraphics[height=6cm, angle=270]{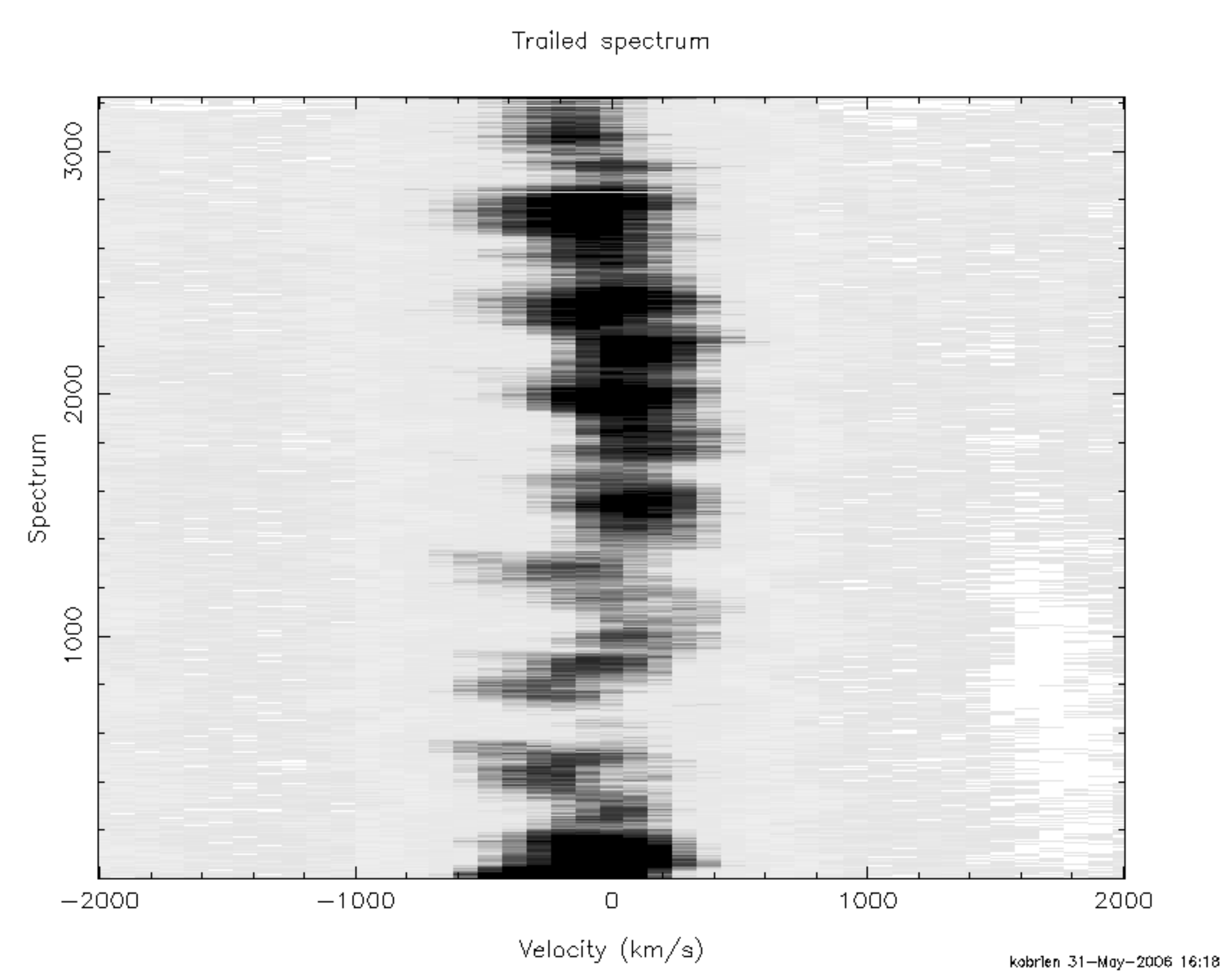} & \includegraphics[height=6cm, angle=270]{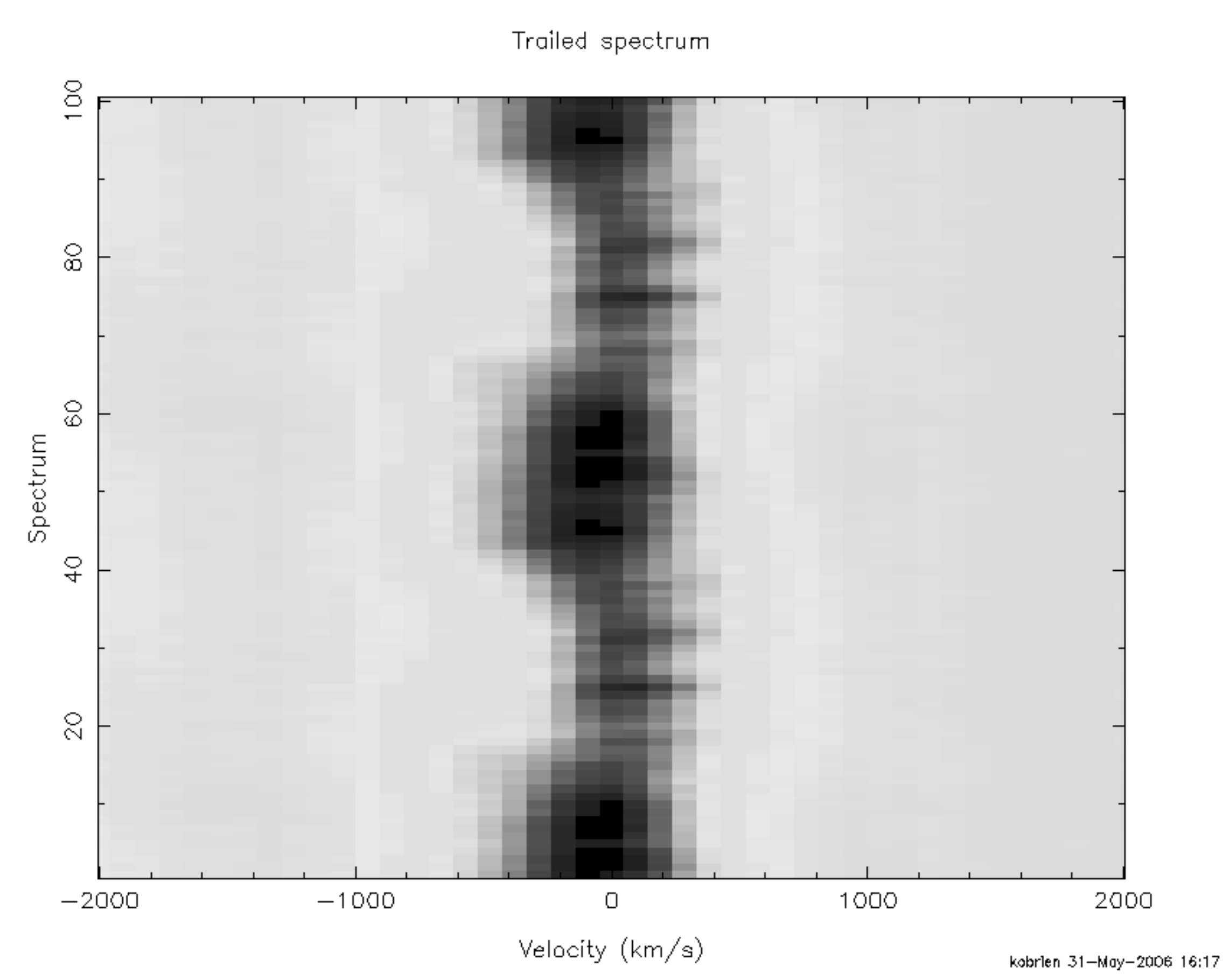} \\
\includegraphics[height=6cm, angle=270]{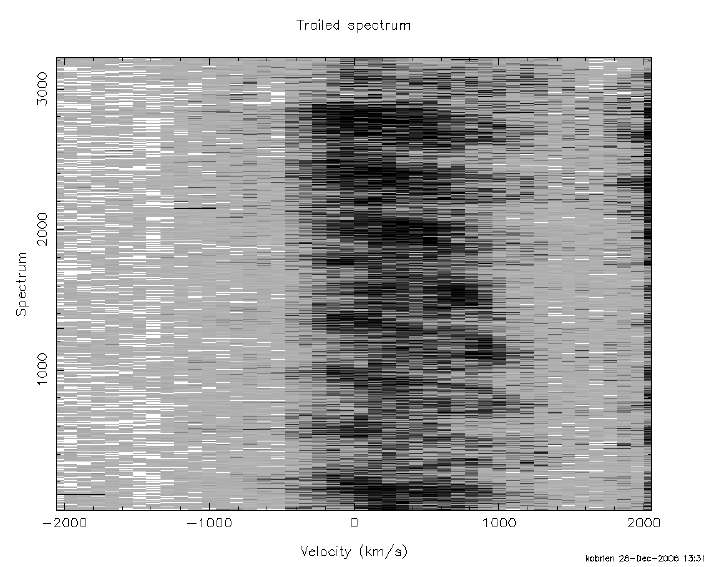} & \includegraphics[height=6cm, angle=270]{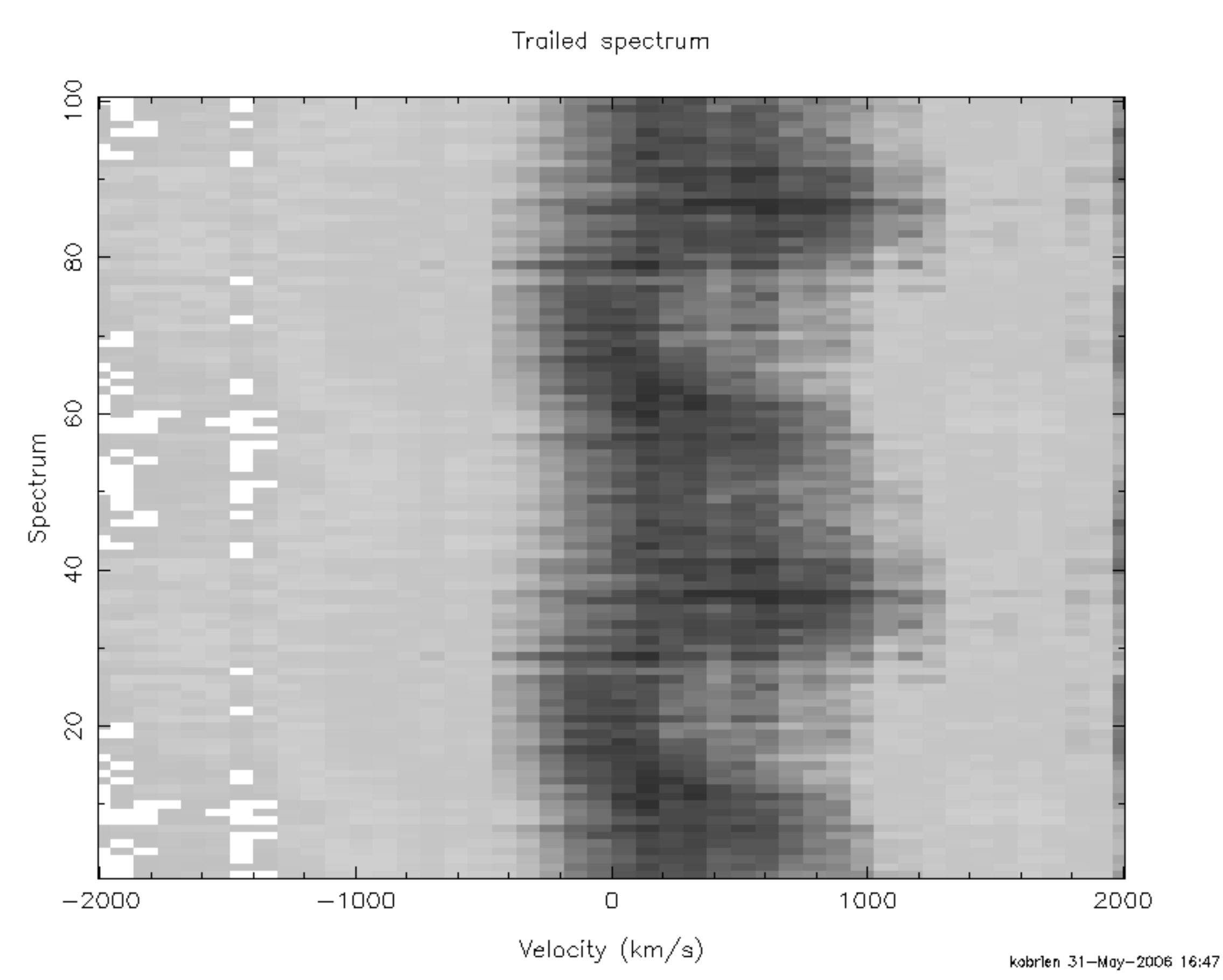} \\
\includegraphics[height=6cm, angle=270]{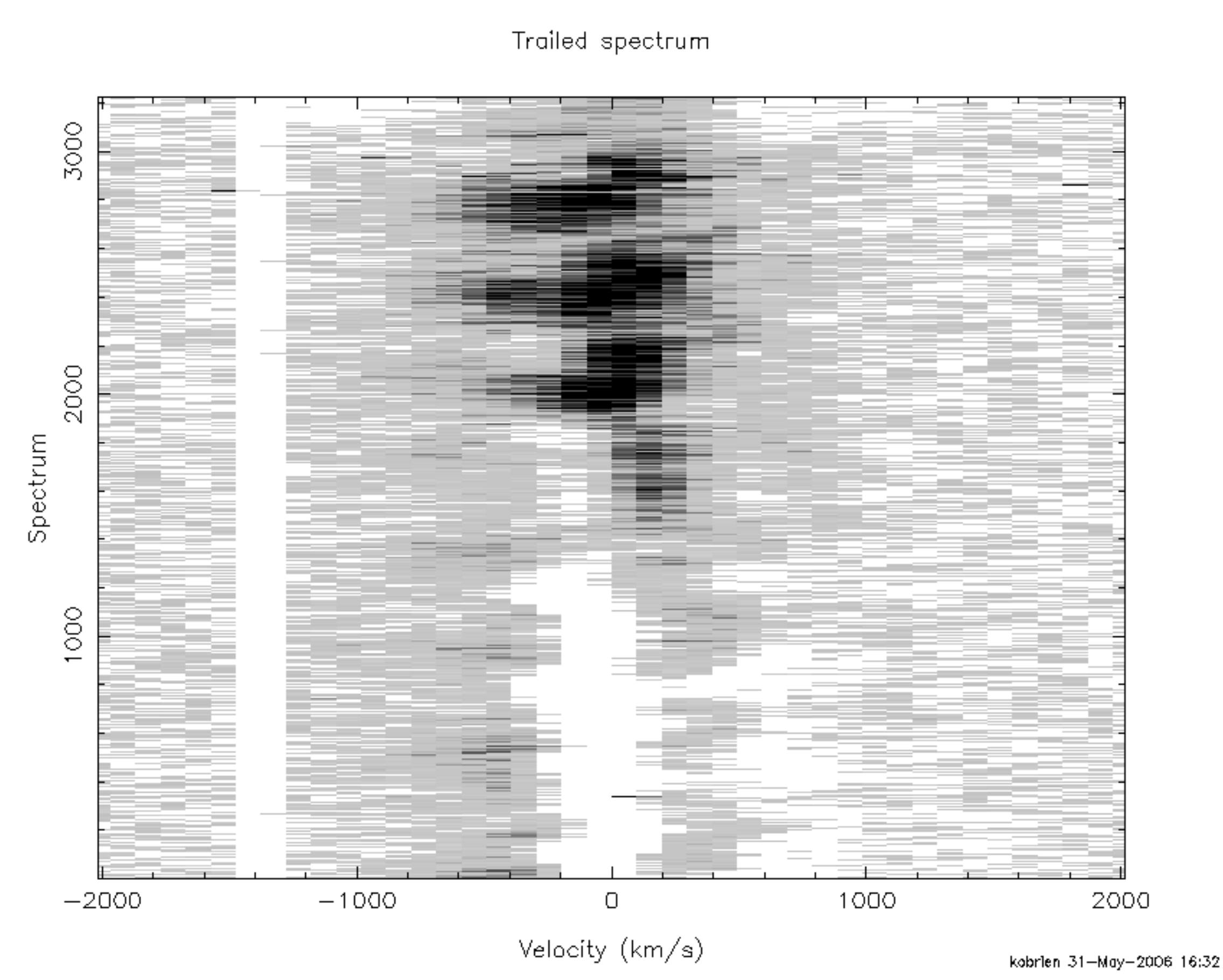} & \includegraphics[height=6cm, angle=270]{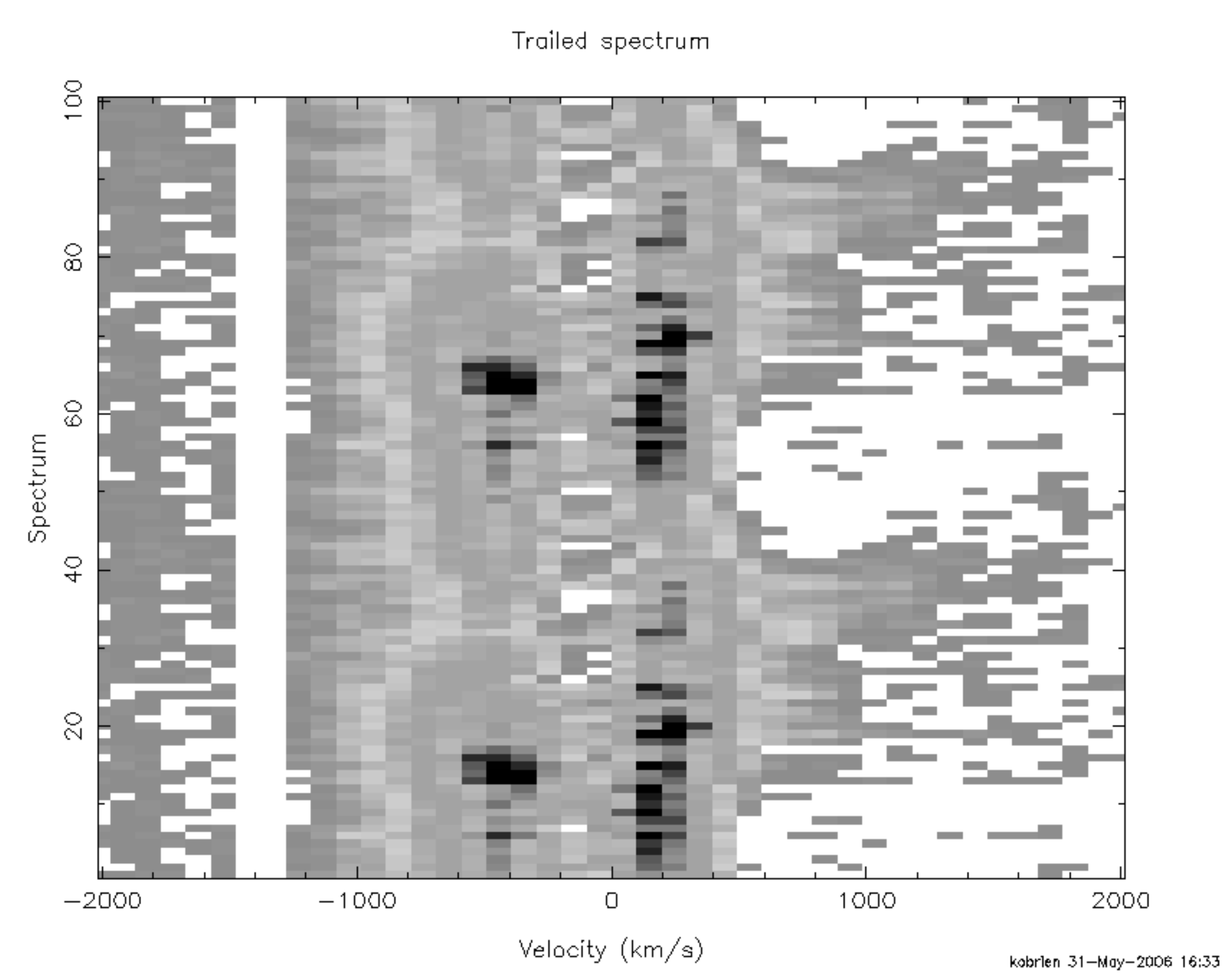} \\
\end{tabular}
\caption{Trailed spectrograms of (from top to bottom) H$\beta$ (486.1~nm), He II (468.6~nm), the bowen blend (464.0~nm) and HeI (447.2~nm). The left-hand panel shows the un-binned trailed spectrogram and the right-hand panel shows the same data folded on the spin period of the white dwarf (the complete spin cycle is repeated for clarity)}
\label{fig:trails}
\end{center}
\end{figure}

As can be seen in Figure~\ref{fig:trails}, there are a number of different types of variability that can be seen in the trailed spectrograms of FO Aqr. In the Balmer series of Hydrogen, clear S-waves can be seen in the phase-binned data which are $180^{\circ}$ out of phase. In the He II ($\lambda 468.6$~nm) and He I ($\lambda 447.2$~nm) trailed spectrograms a number of knots can be seen, as well as transient higher velocity features, possibly originating from the same locations as those described by Marsh, et al. (1996). The bowen blend line profile is interesting and complicated by the multiple components that comprise the blend. I have chosen $\lambda 464.0$~nm as the central wavelength (one of the NII lines), although it is clear that other lines of the blend might be more suitable (e.g. $\lambda 464.7$~nm CIII). Further analysis is needed and is beyond the scope of this work (see e.g. \cite{hellier99}). The primary goal of showing such data is to show the excellent quality that can be obtained. 

\section{Future upgrades}

The HIT mode is now fully operational, it is possible to imagine a number of different ways it could be improved upon. However, it is only possible to make wholesale changes to the instrument on the request of the ESO community, so it is impossible to tell at this stage which ones are likely to be put into effect. Possible upgrades include;

\begin{itemize}
\item A new high-throughput, high dispersion grism more capable of resolving the velocity profiles in Interacting Binaries.
\item An improved time stamping system that has a direct connection to GPS in order to give an independent measure of the arrival time of the photons and remove uncertainty in the accuracy of the current time stamps, which are vital to many of the scientific applications.
\item An EMCCD detector for photon counting spectroscopy and/or continuous readout. This would enable the user to perform fast spectroscopy without the need to stop and read-out the CCD every few tens of frames. This would increase the duty cycle even closer to 100\% and ensure evenly sampled lightcurves.
\item The addition of imaging and/or spectro-polarimetry, which would require a hardware upgrade of FORS2.
\end{itemize}

\section{Conclusion}

The HIT mode of FORS2 is in fact three separate modes, one for imaging and two for spectroscopy. In imaging it is capable of taking images of variable objects with a time resolution in the range 0.6--156~milliseconds through any of the filters of the FORS filter-set. The images are taken through a pseudo-longslit formed by the moveable jaws of the MOS unit and the charge is continuously shifted from the exposed region and stored in the masked region. Once this region is full the CCD is read-out and the process begins again. The only deadtime that occurs is during the read-out of the CCD.

For spectroscopy there are two further modes. The first uses the same clocking scheme as in the imaging mode and hence offers the same time resolution. In addition to this a second mode that operates a `shift-and-wait' scheme, whereby the image of the slit is shifted very rapidly into the masked region of the CCD, thus exposing a new region under the slit. This region remains exposed for a defined amount of time before it too is shifted into the masked region. This process again continues until the masked region is full and the CCD is read out in the usual way. 

The HIT mode is the only mode of its kind available on an 8~m class telescope and as such is a unique capability. I have summarised the issues and potential limitations of the mode as well as show some of the promise of the mode. I hope that in the future the HIT mode will be used in a number of applications and will help to show the strengths of the high time resolution spectroscopy with large telescope both in the 8~m and ELT eras. 

\begin{acknowledgement}
The author would like to thank the many people that have helped and continue to help with the implementation of the HIT mode on Paranal. These include Thomas Szeifert, Nicolas Haddad, Pedro Baksai, Mario Kiekebusch, Claudio Cumani and Karl-Heinz Mantel as well as many other past and present members of the FORS Instrument Operation Team. In addition, I would like to thank Jason Spyromilio and Andreas Kaufer for allowing me to experiment with the instrument. 
\end{acknowledgement}

\end{document}